\documentclass[11pt,psfig,epsfig,graphicx,psfrag]{article}
\usepackage{graphicx}
\usepackage{amsfonts,amssymb,amsmath}
\usepackage{hyperref}

\newcommand{\be}{\begin{equation}}
\newcommand{\ee}{\end{equation}}
\newcommand{\ba}{\begin{eqnarray}}
\newcommand{\ea}{\end{eqnarray}}

\def\L5{\tilde{\Lambda}}

\newcommand{\m}{\hat m}
\def\pd{\partial}
\def\a{\alpha}
\def\b{\beta}

\def\di{\mathrm{d}}

\def\m{\mu}
\def\n{\nu}

\def\p{\pi}

\def\l{\lambda}

\def\r{\rho}
\def\s{\sigma}

\long\def\symbolfootnote[#1]#2{\begingroup%
\def\thefootnote{\fnsymbol{footnote}}\footnote[#1]{#2}\endgroup}

\renewcommand{\d}{{\mathrm{d}}}

\def\1{\mathchoice{\rm 1\mskip-4.2mu l}{\rm 1\mskip-4.2mu l}%
{\rm 1\mskip-4.6mu l}{\rm 1\mskip-5.2mu l}}

\begin{document}
\vskip 1cm

\begin{center}
{\Large {\bf Gauge Symmetry and Consistent Spin-Two Theories\symbolfootnote[2]{To appear in the Proceedings of IRGAC'06, Barcelona 2006.}}}\\[1cm]
D.~Blas$^{a,b}$\footnote{dblas@ffn.ub.es},\\
$^a${\it Departament de F\'\i sica Fonamental, Universitat de Barcelona,\\
Diagonal 647, 08028 Barcelona, Spain.}\\
$^b${\it Center for Cosmology and Particle Physics,\\
Physics Department, New York University, NY 10003, USA.}\\

\end{center}
\vskip 0.2cm

\noindent
\begin{abstract}
We study Lagrangians with the minimal amount of gauge symmetry
required to propagate spin-two particles without ghosts or tachyons.
In general, these Lagrangians also have a scalar mode in their
spectrum. We find that, in two cases, the symmetry can be enhanced to
a larger group: the whole group of diffeomorphisms or a enhancement
involving a Weyl symmetry. We consider the non-linear completions of these
theories. The intuitive completions yield the usual scalar-tensor
 theories except for the pure spin-two cases,
which correspond to two inequivalent Lagrangians giving rise to
 Einstein's equations. A more constructive self-consistent approach
 yields a background dependent Lagrangian.\\

PACS numbers: 04.50.+h,11.15.-q.
\end{abstract}



\section{Introduction}

It has long been known that the minimal amount of gauge symmetry
required for building the Hilbert space of massless spin-two
particles from tensorial objects are the linear ``transverse''
diffeomorphisms \cite{vanderBij} which we will call TDiff (see also
\cite{Weinberg:1964ew,Duff:1974hb} for previous related work) and
which are given by the transformation \be \delta
h_{\m\n}=2\partial_{(\m}\xi_{\n)}, \quad \partial_\m \xi^\m=0. \ee
 The
basic reason for that is that the trace $h=\eta^{\m\n}h_{\m\n}$ is
Lorentz invariant and thus we can restrict to gauge symmetries which
leave it invariant and just act on the spin-one part of the
symmetric tensor \cite{vanderBij,Alvarez:2006uu}. In the first part
of this note we will study the general Lagrangians meeting the
requirement of
 TDiff gauge invariance.

Besides, it is usually claimed that from the consistent self-interaction
 of spin-two particle,
  the whole  group of diffeomorphisms (Diff) is obtained \cite{Ogiev,Deser:1969wk,Boulware:1974sr,Wald:1986bj}.
These approaches are based on the imposition of the whole Diff at linear level.
It is then interesting to study how these
results are modified for the TDiff case. We will be concerned with this in
the second part of the note.

\section{Lorentz Invariant Healthy Lagrangians}

We will first study the ghost and tachyon free Lagrangians for a
symmetric tensor $h_{\m\n}$ that incorporate the TDiff gauge
symmetry. The most general second order Lagrangian for this field is
\begin{eqnarray}
\label{lagr}
       {\mathcal{L}}&=&\frac{1}{4}\partial_\mu
       h^{\nu\rho}\partial^\mu h_{\nu\rho}-\frac{\beta}{2}
       \partial_\mu h^{\mu\rho}\partial_\nu h^\nu_\rho+
       \frac{a}{2}\partial^\mu h\partial^\rho h_{\mu\rho}
       -\frac{b}{4}\partial_\mu h\partial^\mu h-
       \frac{1}{2}{m}^2 \left(h^2-\alpha h_{\m\n}h^{\m\n}\right),
\end{eqnarray}
where the first term is mandatory to propagate spin-two polarizations and
indexes have been manipulated with the Minkowski metric $\eta_{\m\n}$. Let us
consider a general
gauge transformation
\be
\label{gauge}
\delta h_{\m\n}=\tau_{\m\n}+\frac{1}{n}\phi \eta_{\m\n}
\ee
where $\tau^\m_{\phantom{\m}\m}=0$ and $n$ is the dimension of the space-time.
One can prove (cfr. \cite{Alvarez:2006uu})
that this Lagrangian can have a gauge
symmetry only for the case $\beta=1$ and
\be
\tau_{\m\n}=\partial_{(\m}\zeta_{\n)}-\frac{1}{n}\partial_\a \zeta^\a \eta_{\m\n}.
\ee
After imposing $\b=1$, the different choices of parameters in (\ref{lagr}) which give rise to a gauge symmetry, together with the conditions on the parameters of the
gauge transformation (\ref{gauge}) are:\\

\begin{tabular}{l l l}
$\bullet$ TDiff: &$\quad \alpha=0$;& $\quad \phi=0$, $\partial_\r \zeta^\r=0$.\\
$\bullet$ Weyl: &$\quad \alpha=n$, $a=\frac{2}{n}, \ b=\frac{n+2}{n^2}$;& $\quad \zeta_\r=0$.\\
$\bullet$ Diff: &$\quad m^2=0$, $a=b=1$;& $\quad \phi=\partial_\r \zeta^\r$.\\
$\bullet$ Weyl and TDiff (WTDiff): &$\quad m^2=0$, $a=\frac{2}{n}, \ b=\frac{n+2}{n^2}$; &$\quad \partial_\r \zeta^\r=0$.\\
\end{tabular}\\

The previous parameters are unique up to for field redefinition $h_{\m\n}\mapsto h_{\m\n}+\lambda h \eta_{\m\n}$.
 We see that the TDiff invariance requirement implies  $\b=1$ and
$\a=0$. The first of these conditions is necessary also from the direct analysis of
the propagating fields in the theory. To show this, we first decompose the tensor $h_{\m\n}$ into irreducible representations
under the $SO(3)$ subgroup of the Lorentz group,
\ba
h_{00}=A, \quad h_{0i}=\partial_iB+ V_i,\quad
h_{ij}=\psi\delta_{ij}+\partial_i \partial_j E+2\partial_{(i}F_{j)}+t_{ij},
\ea
where $\partial^iF_i=\partial^iV_i=\partial^it_{ij}=t^{i}_{\phantom{i}i}=0$. At the linear level
the different representations decouple, and thus we can study each of them independently. Studying the
vector degrees of freedom $(V_i, F_i)$ we realize that their lagrangian can be expressed as
\be
{\mathcal L}_v=-\frac{\Delta}{2}\left(V_i-\partial_0F_i\right)^2+\frac{1}{2}(\b-1)\left(\partial_0V_i-\Delta F_i\right)^2,
\ee
where $\Delta=\pd_i \pd_i$. This lagrangian has a ghost unless $\b=1$ \cite{Alvarez:2006uu}. We restrict our study to this
kind of Lagrangians which, as we saw before, are TDiff gauge invariant.

Let us first study the massless case $m^2=0$. The Lagrangian for the tensor modes $t_{ij}$ is simply
\be
{\mathcal L}_t=-\frac{1}{4}t^{ij}\square t_{ij}.
\ee
The field $V_i$ is not dynamical and it gives rise to a constraint which also cancels $F_i$, making the vector sector trivial.
The scalar sector $(A,B,\psi,E)$ is more interesting. The field $B$ is a Lagrange multiplier whose variation
produces the constraint
\be
(n-2)\psi=(a-1)h,
\ee
which once substituted back in the Lagrangian gives rise to
\be
{\mathcal L}_s=\frac{C}{4}h\square h
\ee
where $C=b-\frac{1-2a+(n-1)a^2}{n-2}$. Thus, the extra degree of freedom in the theory cancels whenever $C=0$ and
it is well behaved for $C<0$. This means, that from the ghost and tachyon free condition we do not \emph{only}
recover the massless Fierz-Pauli Lagrangian,
but rather
 a perfectly well-defined family of Lagrangians which propagate a spin-two particle and a scalar. If we want to restrict ourselves
to pure spin-two
we find two possibilities which coincide with the choices of enhanced gauge symmetries (see
the previous page):
\begin{itemize}
\item $a=b=1$ and field redefinitions $h_{\m\n}\mapsto h_{\m\n}+\lambda h \eta$,
leading to the whole Diff group \cite{van}.
\item $a=\frac{2}{n}, \ b=\frac{n+2}{n^2}$ which lead to the Weyl and TDiff gauge symmetry (WTDiff)
\footnote{It may seem that
we recover this possibility for $\lambda=-\frac{1}{n}$ in the previous transformation, but notice that in this
case the transformation is singular.}.
\end{itemize}

For the massive case, a similar analysis yields the Fierz-Pauli massive Lagrangian $\b=a=b=\a=1$ as the only possibility.

\section{Non-linear Completions}

From the \emph{strong equivalence principle}, gravity must couple to any kind
of energy including its own \cite{Will:2001mx}. Thus, if the graviton is described by a
 spin-two  particle, this particle must be coupled to its own energy-momentum tensor. If we
do this at linear level, namely if we write the energy-momentum tensor of
the graviton as the source for its equations of motion, this system of equations is no longer
derivable from a Lagrangian and we need to write more non-linear terms. This
process goes on (see \cite{Ortin:2004ms} and references therein) and it is usually stated
that the only solution to this non-linear series is General Relativity with
the usual Einstein-Hilbert action.

However, as we highlighted in the introduction, most of these approaches depart from
the massless Fierz-Pauli Lagrangian (see however \cite{vanderBij}).
As we saw in the previous section, we could consider any of the well-behaved Lagrangians
at the linear level as our starting point and try to find its non-linear completion.
We can do it intuitively \cite{vanderBij,Alvarez:2006uu} and also more constructively (see below).

\subsection{Intuitive Completion}

A possible non-linear extension of the linear TDiff is provided by any subgroup of
the non-linear Diff for which an object $f$, which at the linear level reduces to the trace $h$, transforms as
a scalar\footnote{We restrict to this possibility even if more general transformations could arise.
A constructive way of finding this transformations will be discussed later. Also we restrict
to those objects $f$ of the form (\ref{f}) for the Minkowski metric, but another background
could be chosen.}. That is, given
\be
\label{f}
f\left(\eta_{\m\n},g_{\m\n}\right)=k+\eta^{\m\n}h_{\m\n} + O\left(h_{\m\n}^2\right)
\ee
for $k$ a constant and $h_{\m\n}=g_{\m\n}-\eta_{\m\n}$,
we want to find the subgroup of Diff such that
\be
\delta_\xi f=\xi^\m \partial_\m f,
\ee
for $\delta_\xi g_{\m\n}=2\nabla_{(\m}\xi_{\n)}$. Clearly this subgroup, if it exists, will be
background dependent. The previous condition can be expressed as
\be
\label{condition}
A^\m _\r \nabla_{\m}\xi^{\r}-\xi^\r \partial_\r f =A^\m _\r \partial_{\m}\xi^{\r}=0,
\ee
where
$$A_{\r}^\m=2\frac{\delta f}{\delta g_{\m\n}}g_{\n\r}.$$
In particular this means that the translations are always a subgroup.\\

Let us study the group structure for a generic $f$. From Frobenius theorem applied to the
Diff, the infinitesimal transformations
will be integrable iff \cite{Wald:1986bj}
\be
[\xi_1^\m\partial_\m,\xi_2^\n\partial_\n]=\xi_3^\n\partial_\n
\ee
with $\xi_3^\n=\xi_1^\m\partial_\m\xi_2^\n-\xi_2^\m\partial_\m\xi_1^\n$.
The integrability condition is just
\be
A^\m _\r \partial_{\m}\xi_3^{\r}=2A^\m _\r\left(\partial_\m \xi_{[1|}^\a \partial_\a \xi_{|2]}^\r+
\xi_{[1|}^\a\partial_\m  \partial_\a \xi_{|2]}^\r\right)=0,
\ee
for $\xi_1$ and $\xi_2$ satisfying (\ref{condition}). For the term involving second derivatives to cancel,
the only possibility is  $A_{\r}^\m=l(x)S^\r_\m$, with $S_\r^\m$ being a constant matrix, i.e.
\be
2\delta f=l(x)g^{\m\n}\delta g_{\m\n}=l(x)g^{-1}\delta g,
\ee
where $g=\det g_{\m\n}$. Thus, $f$ depends just on the determinant of the metric.
The subgroup which preserves this functions will be TDiff also at the
non-linear level,
\be
\partial_{\m}\xi^\m=0.
\ee
Once integrated, this subgroup gives rise to the diffeomorphisms of Jacobian equal to one, which
are related to unimodular gravity \cite{vanderBij}.

Let us
consider the simplest function $f=|g|$. We know that
\be
|g|=1+\eta^{\m\n}h_{\m\n}+O(h_{\m\n}^2),
\ee
which in fact holds for any background. General Lagrangians where $|g|$ is considered as
an independent degree of freedom have been
studied in \cite{vanderBij,Alvarez:2006uu} and they are
usually equivalent to scalar-tensor theories of
gravity except for an integration constant.
It is interesting to note that once the Weyl symmetry
\be
\label{conformal}
\delta g_{\m\n}=e^\phi g_{\m\n}
\ee
is also promoted as a gauge symmetry,
we find a unique Lagrangian\footnote{Notice that this Lagrangian can not be put
in the Einstein frame, as it is Weyl invariant.}
\be
\label{WTdiffL}
{\mathcal S}_{WTDiff}=\int \di^4x \hat g^{\m\n} R_{\m\n}(\hat g_{\m\n})+ S_M(g,\hat g_{\m\n},\psi).
\ee
where $\hat g_{\m\n}=|g|^{-1/n}g_{\m\n}$ and $S_M$ refers to a matter Lagrangian compatible with the Weyl symmetry.
This Lagrangian yields Einstein's equations of motion in the gauge $|g|=1$
 (even when coupled to matter) except for the origin of the cosmological constant which comes from
 an integration constant \cite{Alvarez:2006uu}.

Notice also that (\ref{conformal}) could be considered as too restrictive, as what we seek is a transformation of
the determinant of the form
\be
\delta_{(\phi,\xi)} g=\phi g+\xi^\m\partial_\m g.
\ee
However, from the previous expression we find that
\be
[\delta_{(\phi_1,\xi_1)},\delta_{(\phi_1,\xi_1)}]=\delta_{(\xi_{[1}\partial\phi_{2]},\xi_3)}.
\ee
If we want the same algebra to hold for the metric field $g_{\m\n}$ then it is clear that the transformation of
the whole metric must be the usual conformal rescaling, i.e.
\be
\delta_{(\phi,\xi)} g_{\m\n}=\phi^{1/n} g_{\m\n}+2\nabla_{(\m}\xi_{\n)}.
\ee

\subsection{Constructive Completion}

There are different ways in which the non-linear completion can be found constructively. The
most direct one is to consider
the energy-momentum tensor of the graviton as a source for the equations
of motion of the graviton. This amounts to the
first correction, or three-graviton vertex, for the linear action and is not consistent as there is
no Lagrangian that gives rise to these equations of motion
\cite{Ogiev,Ortin:2004ms}. Another way of performing the completion is to first show how the gauge symmetry can
be enlarged non-linearly \cite{Ogiev,Wald:1986bj,Boulanger:2000rq} and then building a Lagrangian endowed
with the non-linear gauge symmetry up to the desired order.  For the case of linearized Diff
 symmetry this non-linear
deformations were first addressed in \cite{Ogiev} and later in \cite{Wald:1986bj}
and \cite{Boulanger:2000rq}.
The equivalent calculation for TDiff and WTDiff is quite cumbersome and will be presented elsewhere \cite{Blas}.

An alternative approach for the Diff case which extends easily to the WTDiff case exists.
This approach is based on the first order formulation of gravity \cite{Deser:1969wk}.
The second order action for the first order formulation of the Lagrangian (\ref{WTdiffL}) is
\be
\label{second}
S^{(1)}=\int \d^n x\left\{-\hat h^{\m\n}2\partial_{[\m}\Gamma_{\r]\n}^{\phantom{\r|\n}\r}+\eta^{\m\n} 2
\Gamma_{\l[\m}^{\phantom{\p|\n}\r}\Gamma_{\r]\n}^{\phantom{\p|\n}\l}\right\}
\ee
where $\hat h_{\m\n}=h_{\m\n}-h\eta_{\m\n}$ and the metric and the connection
are now considered as independent fields. The equations of motion from the variation of $\hat h_{\m\n}$ are the traceless part
of the Fierz-Pauli case, whereas from the variation of $\Gamma_{\m\n}^{\phantom{\r|\n}\r}$ we find a
 constraint for this field which, once solved,
yields (for $n\neq 2$)
\be
\Gamma_{\m\n}^{\phantom{\r|\n}\r}=\frac{1}{2}\eta^{\r\s}\left(\pd_\m \hat h_{\n\s}+\pd_\n \hat h_{\m\s}-\pd_\s \hat h_{\m\n}\right).
\ee
This is just the equation of compatibility of the connection and the traceless metric at linear order. Substituting this constraint
in the Lagrangian we just get the WTDiff Lagrangian for $h_{\m\n}$. To calculate the energy-momentum tensor we use
the Rosenfeld prescription for which we need to assign a weight to the fields $\hat h_{\m\n}$
and $\Gamma_{\m\n}^{\phantom{\n\m}\r}$
which is the strongest assumption of Deser's method \cite{Ortin:2004ms}. If we consider $\hat h^{\m\n}$ to be
a tensor density and the indices of the connection to be vectorial, it is easy to see that
the energy-momentum tensor is
 given by
the usual energy-momentum tensor
of \cite{Deser:1969wk} except for the fact that the tensor $\hat h_{\m\n}$ is now traceless.
 The WTDiff gauge symmetry implies that the object to couple to the free equations
 of motion of (\ref{second}) is the traceless part of the energy-momentum tensor. Following \cite{Deser:1969wk},
this coupling can be derived from the term
\be
\label{nlcompl}
{\mathcal S}^{(2)}=-2\int \di^n x \hat h^{\m\n}\Gamma_{\r[\m}^{\phantom{\n)\r}\s}
\Gamma_{\s]\n}^{\phantom{\n)\r}\r}.
\ee
as $\hat h^{\m\n}$ is already traceless.
Thus the action at third order simply reads:
\be
\label{nllagr}
{\mathcal S}\equiv {\mathcal S}^{(1)}+{\mathcal S}^{(2)}=
\int \d^n x \tilde g^{\m\n} R_{\m\n}\left(\Gamma_{\a\b}^{\phantom{\m\n}\r}\right),
\ee
where $\tilde g^{\m\n}=\eta^{\m\n}-\hat h^{\m\n}$.
 This Lagrangian differs from the one which we guessed intuitively and is background dependent as $\hat h_{\m\n}$ involves
$\eta_{\m\n}$ in its definition\footnote{Note that
the non-linear TDiff of the previous subsection depend only on a volume form.}. Besides, the equations of motion are not Einstein's equations but rather
\be
\label{eomc}
R[\tilde g]_{\m\n}-\frac{1}{n}\eta_{\m\n}R[\tilde g]=0
\ee
where the connection is compatible with the metric associated to the density tensor $\tilde g^{\m\n}$,
 $g_{\m\n}$ which satisfies the constraint
\be
\label{cond}
\sqrt{-g}g^{\m\n}\eta_{\m\n}=n.
\ee
This condition is preserved by diffeomorphisms satisfying
\be
\eta_{\m\n}\left(g^{\m\a}\delta^{\n}_{\b}-\frac{1}{2}\delta^{\a}_{\b}g^{\m\n}\right)\nabla_{\a}\xi^{\b}=0,
\ee
which reduces to the transverse condition at the linear level. Again, the algebra of these diffeomorphisms
does not close for general metrics and thus  do not constitute a subgroup. Even vacuum solutions
for the equations (\ref{eomc}) differ from Einstein's equations and we leave for future work
the actual computation of phenomenological constraints of the theory which appear at the non-linear
level \cite{Blas}.

The reason why we have not
found the WTDiff Lagrangian at the non-linear level is the highly non-linear dependence
of the determinant of the metric as expressed in terms of traces. Remember that for the
 non-linear WTDiff
the action is expressed in terms of a tensor with $\hat g=1$. It is impossible to
find this condition for the determinant from the linear conditions on the trace with respect to the
metric $\eta_{\m\n}$ in one single step. However,
considering the field $\hat h_{\m\n}$ as a tensor density, the arguments of \cite{Deser:1969wk}
apply also here and we are directly selecting the condition (\ref{cond})
as the non-linear one. \\

\section{Conclusions}

In this note we have shown that the requirement of ghost and tachyon free
Lagrangians which describe spin-two particles is satisfied by a whole family
of Lagrangians at the linear level which satisfy a reducible gauge symmetry,
TDiff.
 In all the cases but two there is a scalar mode which propagates
 whose mass is constrained by usual phenomenological bounds \cite{Alvarez:2006uu,Will:2001mx}.
The two special cases are the usual Diff case, where the theory becomes
irreducible and WTDiff, which incorporates a Weyl gauge transformation.\\

We have seen that the intuitive non-linear completion of these theories gives
rise to Lagrangians which differ from the usual Einstein-Hilbert Lagrangian
but which reduce again to scalar-tensor theories of gravity except
for the previous two cases. In these cases, the equations of motion are Einstein's equations in
a certain gauge except for the origin of the cosmological constant as happens in unimodular
gravity \cite{Weinberg:1988cp,Alvarez:2005iy}. A more constructive approach to the non-linear theory
is possible in two ways.
First, we can work
in the first order formalism of gravity and consistently couple the
conserved energy-momentum tensor of the free action as a source of the equations of motion for the
graviton itself.
The standard derivation of \cite{Deser:1969wk} holds  except for the fact that the metric
must satisfy the extra constraint (\ref{cond}). This way we find a consistently-coupled Lagrangian (\ref{nllagr})
 which
 does not produce Einstein's equations at the non-linear level and is background
 dependent. This illustrates
the non-uniqueness of the derivation in \cite{Deser:1969wk}.  It is not clear whether
other  choices of linear variables or
of weights for $\hat h^{\m\n}$ and $\Gamma_{\phantom{\m}\r\s}^{\m}$ exist that
can reproduce the lagrangian (\ref{WTdiffL}) using similar methods.

The other possibility is to deform the linear algebra
and construct the non-linear group of symmetry. This
approach is currently under investigation \cite{Blas}.\\

I would like to thank E. \'Alvarez, J. Garriga, S. Iblisdir and E. Verdaguer for useful discussions
and for a critical reading of this note. I would also like to thank the CCPP and Physics Department of NYU
for its warm hospitality.


\end{document}